\documentclass{moriond}
\usepackage{amsmath,amssymb}

\def\Journal#1#2#3#4{{#1} {\bf #2}, #3 (#4)}

\def\PRD{{\em Phys. Rev.} D}

\def\be{\begin{equation}}
\def\ee{\end{equation}}
\def\bea{\begin{eqnarray}}
\def\eea{\end{eqnarray}}

\newcommand\D{\mathcal D}
\def\L{\mathcal{L}}
\def\Lm{\mathcal{L}_m}

\newcommand{\Photo}{}

\begin{document}
\vspace*{4cm}
\title{Standard quantum field theory from entangled relativity}

\author{Olivier Minazzoli}

\address{Artemis, Universit\'e C\^ote d'Azur, CNRS, Observatoire C\^ote d'Azur\\ BP4229, 06304, Nice Cedex 4, France,\\Bureau des Affaires Spatiales, 2 rue du Gabian, 98000  Monaco}

\maketitle\abstracts{
Despite its non-linear form, \textit{entangled relativity} possesses both \textit{general relativity} and standard quantum field theory in a specific (but generic) limit. On one side it means that the theory is consistent with our current understanding of elementary physics. But on the other side it means that our current understanding might actually just be approximately valid: and this, surprisingly, goes for both \textit{general relativity} and standard quantum field theory together.}



As we shall see in this communication, \textit{entangled relativity} is a general theory of relativity that is more economical than \textit{general relativity} coupled to matter fields when embedded in a quantum field theory framework, because it requires only two universal dimensionful parameters to be defined, whereas it has all the same ingredients otherwise.
Indeed, let us start from its path integral:

\be
Z = \int \D g  \prod_i \D f_i \exp \left[-\frac{i}{2 \epsilon^2} \int d^4_g x \frac{\L^2_m(f,g)}{R(g)} \right], \label{eq:ERPI}
\ee
where $\int \D$ relates to the sum over all possible field configurations, 
$R$ is the usual Ricci scalar that is constructed upon the metric tensor $g$, $ \mathrm{d}^4_g x := \sqrt{-|g|}  \mathrm{d}^4 x$ is the spacetime volume element, with $|g|$ the metric $g$ determinant, and $\L_m$ is the Lagrangian density of matter fields $f$---which could be the current \textit{standard model of particle physics} Lagrangian density, but most likely a completion of it. One can check that the dimension of what is historically called ``the action'' turns out to be an energy squared. Thus, the only parameter of the theory is a quantum of energy $\epsilon$. This parameter and the causal structure constant $c$---hidden in the spacetime volume element---are the only two universal constants of the theory.

At this stage, one can already deduce an important fact about this theory: it does not have a quantum of action. As a consequence, one can already deduce that the quantum of action $\hbar$ can only be effective, rather than elementary. Obviously, this changes quite a bit from the XXth century picture of elementary physics.

Another important fact that one can draw at this stage is that one cannot construct a length scale or a time scale from a quantum of energy ($\epsilon$) and a speed ($c$). Hence, right from the beginning, one can deduce that there is no notion of elementary units of space, or of time, in this theory. Given that all the real troubles of quantum \textit{general relativity} are linked, one way or another, to the existence of the Planck length and time \cite{kiefer:2012bk}, it is a rather unexpected and interesting fact about this theory. In particular, it means that there is no reason a priori to expect anything special happening to the smooth structure of spacetime at any scale---unlike what has been realized very early on in quantum \textit{general relativity} \cite{bronstein:1933}. This would be a good news, as one would likely not have to (fundamentally) come up with a discretization recipe for the computation of the path integral Eq. (\ref{eq:ERPI}), such as in \textit{causal dynamical triangulation} or in \textit{Regge calculus} for (non-perturbative) quantum gravity---although such types of discretization procedures may still be necessary approximations in order to be able to actually evaluate the path integral numerically \cite{kiefer:2012bk}. Therefore, quantum \textit{entangled relativity} should be a radically new direction to explore and to evaluate in the field of quantum gravity.

A common na\"ive comment that some may have when confronted for the first time to Eq. (\ref{eq:ERPI}), among many, is simply that it cannot make sense, given the infamous $\Lm^2$ in the nominator. Indeed, if one assumes that gravity can be neglected at the scale of particle physics experiments, then $R$ must be constant---if not much worse, $R=0$---and one ends up with a theory that does not make any sense from a quantum field theory point of view---nor from any point of view for that matter. But the comment is na\"ive, because it assumes that neglecting gravity only implies that the variation of the metric tensor can be neglected at the scale of particle physics. But the classical theory that derives from the extremization of the quantum phase in Eq. (\ref{eq:ERPI}) is not \textit{general relativity}. Hence, one does not know, a priori, what neglecting gravity even means in this context. In order to figure it out, one has no choice than to study the gravitational (classical) phenomenology of the theory.

Fortunately enough, the classical gravitational phenomenology of the theory turns out to correspond to a special case of a class of theories studied some ten years ago \cite{{minazzoli:2013pr}}. Indeed, let us start from the following action
\be \label{eq:pressuron}
S \propto \int d^4_g x \frac{\Phi}{2 \alpha} \left[R(g) - \frac{\omega(\Phi)}{\Phi^2} (\partial \Phi)_g^2 + \frac{2 \alpha}{\sqrt{\Phi}} \Lm(f,g)\right], 
\ee
where $\omega(\Phi)$ is an arbitrary function, and $\alpha$ a normalization constant that is such that $\alpha/\Phi_0 = 8 \pi G /c^4$, where $\Phi_0$ is the background value of the scalar-field in the solar system for instance. The field equations that derive from this action read
\begin{equation}
R^{\mu \nu}= \alpha \frac{1}{\sqrt{\Phi}}\left[T^{\mu \nu}-\frac{1}{2} g^{\mu \nu} T\right]+\frac{1}{\Phi}\left[\nabla^\mu \partial^\nu \Phi+\frac{1}{2} g^{\mu \nu} \Box \Phi\right] +\frac{\omega(\Phi)}{\Phi^2} \partial^\mu \Phi \partial^\nu \Phi,
\end{equation}
with
\begin{equation} \label{eq:eqphi}
\frac{2 \omega(\Phi)+3}{\Phi} \square \Phi+\frac{\omega_{, \Phi}(\Phi)}{\Phi}\left(\partial_\sigma \Phi\right)^2=\alpha \frac{1}{\sqrt{\Phi}}\left[T-\mathcal{L}_m^o\right],
\end{equation}
where $\Lm^o$ is the on-shell value of the matter Lagrangian, and
\begin{equation}
\nabla_\sigma T^{\mu \sigma}=\frac{1}{2}\left(\mathcal{L}_m^o g^{\mu \sigma}-T^{\mu \sigma}\right) \frac{\partial_\sigma \Phi}{\Phi}.
\end{equation}
This theory is well-defined for all $\omega \neq -3/2$, and notably imply the same post-Newtonian parameters as \textit{general relativity}---that are $\gamma = \beta =1$---for all $\omega > -3/2$. It takes literally two lines of calculation to show that the case $\omega = 0$ is equivalent, at the classical level, to the theory that derives from the extremization of the quantum phase in Eq. (\ref{eq:ERPI}), provided that $\Lm \neq \emptyset$ in the action.\footnote{Just inject back the scalar field equation that derives from Eq. (\ref{eq:pressuron}) with $\omega=0$ and $\Lm \neq \emptyset$ in the original action (\ref{eq:pressuron}), and see that this on-shell action simply is $-\alpha/2 ~\Lm^2/R$. Of course, one can check that all the field equations that derive from both actions are the same, but it takes some more work to check it out \cite{ludwig:2015pl}.} 
In the solar system for instance \cite{{minazzoli:2013pr}}, the field equations from Eq. (\ref{eq:pressuron}) are such that the gravitational field $\Phi$ is constant---or, at least, varies much less than the spacetime metric---for all $\omega > -3/2$. This as been named an \textit{intrinsic decoupling} \cite{minazzoli:2013pr}. It turns out that this remains true in general for a universe mainly made of dust and null electromagnetic radiation such as ours. {\bf But what does it mean for the theory in Eq. (\ref{eq:ERPI}) that $\Phi$ varies much less than the spacetime metric?} Well, it simply notably means that the ratio between $\Lm^o$ and $R$ is constant whenever gravity can be neglected.\footnote{If you are bothered by that, keep in mind that it is not that different from the fact that the ratio between $T$ and $R$ is constant in \textit{general relativity}. $\Lm^o$ is just another scalar quantity that is constructed from matter fields, which sometimes even is such that $\Lm^o = T$. $\Lm^o = T$ for dust and electromagnetic radiation for instance. 
} Indeed, at the classical level, one has $\sqrt{\Phi} = - \alpha~\Lm^o /R$ for $\omega =0$, as one can check from the derivation of the field equations. Let us specify our case to the $\omega=0$ one for readability, and rewrite our action as follows
\be
S \propto \int d^4_g x \frac{1}{\kappa}\left(\frac{R(g)}{2 \kappa} + \L_m(f,g) \right), \label{eq:ERGR}
\ee
with $\kappa = \alpha / \sqrt\Phi $, whose solution is such that $\kappa = - R / \Lm^o$. It turns out that this field definition is slightly more general than the previous one, as $\kappa$ can in principle also be negative with this definition; whereas $\sqrt{\Phi}$ could not, by construction. $\kappa <0$ should not happen in the observable universe though. Indeed, $\kappa$ for any local solution (e.g. neutron star, black-hole, solar system etc.) has boundary conditions that are such that $\kappa = 8 \pi G / c^4$ at the boundary, usually corresponding to the constant background of the scalar-field at the local scale. From there, one can check that, even for the ultra relativistic density of neutron stars, $\kappa$ varies a few percent only, at max \cite{arruga:2021pr}. Therefore, it has no chance of ``crossing the line'' for the densities of the celestial bodies that exist in the observable universe, and which are not hidden behind an horizon. Nevertheless, one cannot exclude that $\kappa$ can in principle become negative in even denser situations than neutron stars. Rather than a bug of the theory, I think this might end up being a potential interesting feature of the theory, perhaps for instance for describing the primordial universe and/or the behavior of matter inside black-holes. 


{\bf Fine, but what can we say about quantum field theory when gravity can be neglected in Eq. (\ref{eq:ERPI})?} We can say that the metric field does not vary at the scale of particle physics, as usual, but we can also say that the ratio between $\Lm$ and $R$---which is a gravitational degree of freedom ($\kappa$ or $\Phi$ in Eq. (\ref{eq:eqphi}) with $\omega=0$)---does not vary either. Hence, when one neglects gravity, Eq. (\ref{eq:ERPI}) reduces to

\be
Z \approx \int \prod_i \D f_i \exp \left[\frac{i}{ \kappa \epsilon^2} \int d^4 x \L_m(f) \right]. \label{eq:ERPI_sQFT}
\ee
From this limit, one can now identify the quantum of energy $\epsilon$ that was the only free parameter of the theory in Eq. (\ref{eq:ERPI}). Indeed, in order to match with standard quantum field theory on ``flat spacetime'' \footnote{Let us stress the obvious: as far as standard physics goes, the concept of a flat spacetime is not something that exist in the universe (anywhere), since it is not solution of \textit{general relativity} with a cosmological constant. (Obviously, a flat spacetime is also prohibited in \textit{entangled relativity}). Hence, in standard physics, a ``flat spacetime'' is just an useful approximation on scales at which the gradient of the spacetime metric can be neglected. Nothing more than that. So, quantum field theory on ``flat spacetime" is just quantum field theory when gravity is neglected.}, one must have 
\be
\kappa \epsilon^2 = c \hbar, \label{eq:id}
\ee
such that $\epsilon$ turns out to be the (reduced) Planck energy $\epsilon = \sqrt{c \hbar/\kappa}$. This shows that, when gravity can be neglected, the weird looking non-linear phase in Eq. (\ref{eq:ERPI}) recovers standard quantum field theory on ``flat spacetime''. This is far from being trivial, and it boils down to the gravitational (classical) phenomenology of the theory that implies that $\kappa$ varies much less than the spacetime metric for a universe like ours. 

There is exactly one constant on each side of Eq. (\ref{eq:id}): $\epsilon$ on the left hand side and $c$ on the right hand side. Hence, as expected from the start of this communication, $\hbar$ is not an elementary constant of nature in this framework, but rather emerges as such in some limit of the theory. Indeed, although Eq. (\ref{eq:id}) has been obtained in the $\kappa=$constant limit, the value of $\kappa$ can change a few percent inside a neutron star for instance \cite{arruga:2021pr}. So, even if at the scale of a particle physics phenomenon, $\kappa$ can be approximated to be constant on the relevant particle physics scale, it may change depending on the location of the phenomenon inside, or close to, the neutron star, a few percent over several km. Hence, it means that effectively, $\hbar$ has a value that varies depending on the location. This is a novel prediction, which does not depend on any free theoretical parameter, and that eventually might be probed at the observational or experimental level---although the level of variation of $\hbar$ should be minute in the observable universe due to the \textit{intrinsic decoupling} mentioned above. One can even conjecture that the increase of the apparent fine structure constant (potentially) observed in strong gravitational fields \cite{hu:2021mr} might be related to that. Indeed, what is interpreted as a variation of the fine structure constant ($\alpha_{e}$) could likely be interpreted as a variation of Planck's quantum of action $\hbar$ instead---keeping $\alpha_{e}$ constant. (I thank my friend Aurélien Hees for pointing that out).

{\bf What does that mean for standard quantum field theory?} In standard quantum field theory, the path integral and canonical quantization are two sides of the same thing. Indeed, as Feynman famously first showed for quantum mechanics, a path integral ought to lead to the same results as canonical quantization. Hence, from Eq. (\ref{eq:ERPI_sQFT}), one can expect the usual commutation relation between conjugate quantities, but rewritten as follows
\be
[\hat A,\hat B] \approx i  \frac{\kappa \epsilon^2}{c} \mathbb{I}, \label{eq:com}
\ee
where $\hat A$ and $\hat B$ are two arbitrary canonical conjugate quantities---as long as the spatio-temporal scales of the quantum phenomena considered are small with respect to the spatio-temporal variations of the background value of $\kappa$ in the \textit{semiclassical} limit of the theory. However, this should no longer be correct beyond the $\kappa=$constant limit---that is, when $\kappa$ varies significantly even at the scale of particle physics. But one does not even know how one could write Eq. (\ref{eq:com}) in that situation, given that the very notion of a constant quantum of action disappears. Therefore, one may expect that one looses the equivalence between canonical quantization and the path integral when $\kappa$ can no longer be approximated to be constant at the scale of particle physics---that is, way before the quantum gravity regime. One may even conjecture that the procedure of canonical quantization might therefore only be approximately valid in the $\kappa=$constant limit, but actually not valid at an elementary level. 
This would be quite a blow for the various programs of canonical quantization of gravity, such as \textit{Loop Quantum Gravity}. 
Studies of quantum \textit{entangled relativity} should therefore probably concentrate on the path integral formulation. In any case, because of the intertwined nature of the theory, one can expect that quantum \textit{entangled relativity} will also highly depend on the definition of $\Lm$ that will be considered to be valid up to the Planck energy scale. This should drastically complicate any of its investigation as there is no longer any sense to consider the general theory of relativity and the theory of matter fields individually, and one can expect that only the complete theory---that is, Eq. (\ref{eq:ERPI}) with an appropriate definition of $\Lm$---will make sense at the Planck energy scale.


\end{document}